# On Neutrino Masses and Leptonic Mixing

Paul M. Fishbane
Physics Dept. and Institute for Nuclear and Particle Physics,
Univ. of Virginia, Charlottesville, VA 22901

Peter Kaus
Physics Dept., Univ. of California, Riverside, CA 92521

Abstract

Using recent data on neutrino oscillations, we argue that a hierarchical solution for neutrino masses in a three family context is possible, and that the masses of the $\tau$ and $\mu$ neutrinos are very nearly determined within that possibility. We also examine the predictions of a model that determines neutrino and charged lepton mass matrices as well as its consistency with data.

# 1. Introduction

The purpose of this note is to recognize some simple features of neutrino masses within the context of the most conservative assumptions, and to test these features within an interesting predictive model of family symmetry. Our conception of conservative is the following:

1. Neutrinos come in three families, with no additional species, sterile or otherwise.

2. Neutrinos come in right-handed singlets and left-handed doublets, consistent with the other fermions of the standard model. They have mass either by a Dirac term alone in the Lagrangian or by a combination of Dirac terms and Majorana terms that combine through a seesaw mechanism.

3. Like the other sets of fermion families, the masses of neutrinos are hierarchical, with the $v_\tau$ having the largest and the $v_e$ having the smallest mass. (The data to be discussed below are completely consistent with this assumption.)

4. The structure of the mass matrices of the charged leptons and the neutrinos determine neutrino oscillations of the type that can explain recent experiments.

5. The overall scales of the masses in the different fermion sectors differ, and we have nothing special to say about these scales (although we recall the grand unified theory relation $m_b = m_\tau$, a relation that follows in the model we discuss in Section 3). Instead, we are interested only in mass ratios within a given family; the heaviest mass family of a sector will be normalized throughout to unity.

    These assumptions derive essentially from the idea that neutrinos are "normal," that is, they have a hierarchical three-family structure similar to that of the other three sets of fermions of the standard model, the up quarks, the down quarks, and the charged leptons[1]. In each case, the masses of the fermions can be written as an overall scale times powers of the sine of the Cabibbo angle—equivalently the Wolfenstein expansion parameter—$\lambda$. But this fact is not very limiting—we have not yet specified the quantitative nature of the hierarchy. We shall describe in Section 2 the freedom we think this gives us for the neutrino sector, as well as the limits set by the data. The masses of the $v_\mu$ and $v_\tau$ are very nearly predicted.

    A recent set of models [1,2] suggest within the framework of a low-energy supersymmetric extension of the standard model that the existence of mass hierarchies *within* fermionic sectors imply at least one additional U(1) family symmetry, one of which must be anomalous, with a cancellation of its anomaly through the Green-Schwarz mechanism then implying the presence of relations *across* fermionic sectors [3]. This has the additional benefit of predicting $\lambda$ itself. Effectively, the standard model is cut off at the grand unified scale, and couplings are suppressed by powers of U(1) charges. These

---

[1] We should note here that in contrast to the other fermions, neutrinos can have Majorana mass terms and hence a seesaw mechanism; this is an argument for neutrinos not being "normal." We simply assume that in spite of what may be different dynamics neutrinos look like the other fermions.

suppressions appear in the mass matrices when symmetry is broken at the GUT scale, and correspond to the powers of $\lambda$ that appear in these matrices. Such models are especially interesting because the additional U(1)'s are characteristic of superstrings, and moreover, these symmetries are broken by effects associated with strings.

The mass matrices of both the quarks and the fermions—not to mention the relations across families and the value of the Cabibbo angle itself—are thereby predicted. One has not only the charged and neutral leptons mass eigenvalues but the leptonic analog of the Cabibbo-Kobayashi-Maskawa (CKM) mixing matrix[2] as well. With the generic form predicted in this model the mass hierarchies predicted for the leptons cannot be accommodated within the framework of the current data; however, with perturbative extensions it is possible to do so. And while the mixing matrix itself cannot take a form consistent with the maximal mixing between the $\nu_\mu$ and $\nu_\tau$ sectors, it is possible to come reasonably close to this limit. As we shall see, the leading mixing terms between the $\nu_e$ and the other neutrinos is then determined, suggesting ways to test the model further.

## 2. Remarks on Neutrino Masses

Recent experiments [4] have provided quantitative results on neutrino mass differences.

$$\Delta_{\tau\mu} = m_{\nu\tau}^2 - m_{\nu\mu}^2 \text{ and } \Delta_{\mu e} = m_{\nu\mu}^2 - m_{\nu e}^2. \tag{2.1}$$

The remaining available atmospheric and long baseline data [5], if not the short baseline accelerator data [6], is consistent (for recent analyses see [7] and [8]) with what we refer to in Section 1 as the conservative point of view. This data is consistent with the small angle MSW explanation [9] of the solar neutrino data [10] and the large $\nu_\mu$-$\nu_\tau$ mixing explanation of the atmospheric neutrino anomaly. For recent discussion of how one could accommodate all the current data, including that of ref. [6] within a nonconservative framework, we refer the reader elsewhere [11].

In order to ask what the quantitative nature of the hierarchy is, we first discuss the charged leptons, for which we know the numbers. (While the mass scale at which we discuss the quarks is important, for the leptons this is not an issue.) Are the charged leptons like the quarks? The charged leptons seem to steer a course between the two quark families. The up quarks have a mass hierarchy of the form $1:\lambda^4:\lambda^8$ and the masses of the down quarks have the ratio $1:\lambda^2:\lambda^4$. The charged leptons have a mass hierarchy of the form $1:\lambda^2:\lambda^6$. In other words, $\mu/\tau$ is like $s/b$ but $e/\mu$ is like $u/c$. To find out what the data can tell us we define the (for now) strictly phenomenological parameter $R$ by

$$\mu_{\nu e}/\mu_{\nu\mu} = R\, \mu_{\nu\mu}/\mu_{\nu\tau} \tag{2.2}$$

This parameter has the advantage of emphasizing the role of mass ratios, as is appropriate for a hierarchy. The analog to this quantity for the down quarks is very close to unity, and if for the up quarks one replaces the masses by their squares, it is also close to unity. Alternatively, the charged leptons are associated with an analog to $R$ of $O(\lambda^2)$.

---

[2] Ramond et al refer to this as the MNS matrix after the early work of Z. Maki, M. Nakagawa, and S. Sakata, Prog. Theo. Phys. **28**, 247 (1962).

We can regard Eqs. (2.1) and (2.2) as three equations for the three neutrino masses in terms of the three parameters $\Delta_{\tau\mu}$, $\Delta_{\mu e}$, and $R$. We take the first two parameters as given at the values $\Delta_{\tau\mu} = 10^{-3}$ eV$^2$ and $\Delta_{\mu e} = 5 \times 10^{-6}$ eV$^2$—we include the effects of errors at the end of this section—and plot the results for the masses (or their ratios) as a function of $R$. Because there are mass squares present, the resolution of the equations involves four branches, two of which involve only sign difference and are of no importance. Two of the branches give complex values for $R$ is some critical range and the other two give complex values for $R$ outside a related critical range. The precise range limits depend on the $\Delta$'s and have no special importance. Ignoring sign questions, we can reduce the solutions for the masses to two branches, as shown in Figs. 1 to 3.

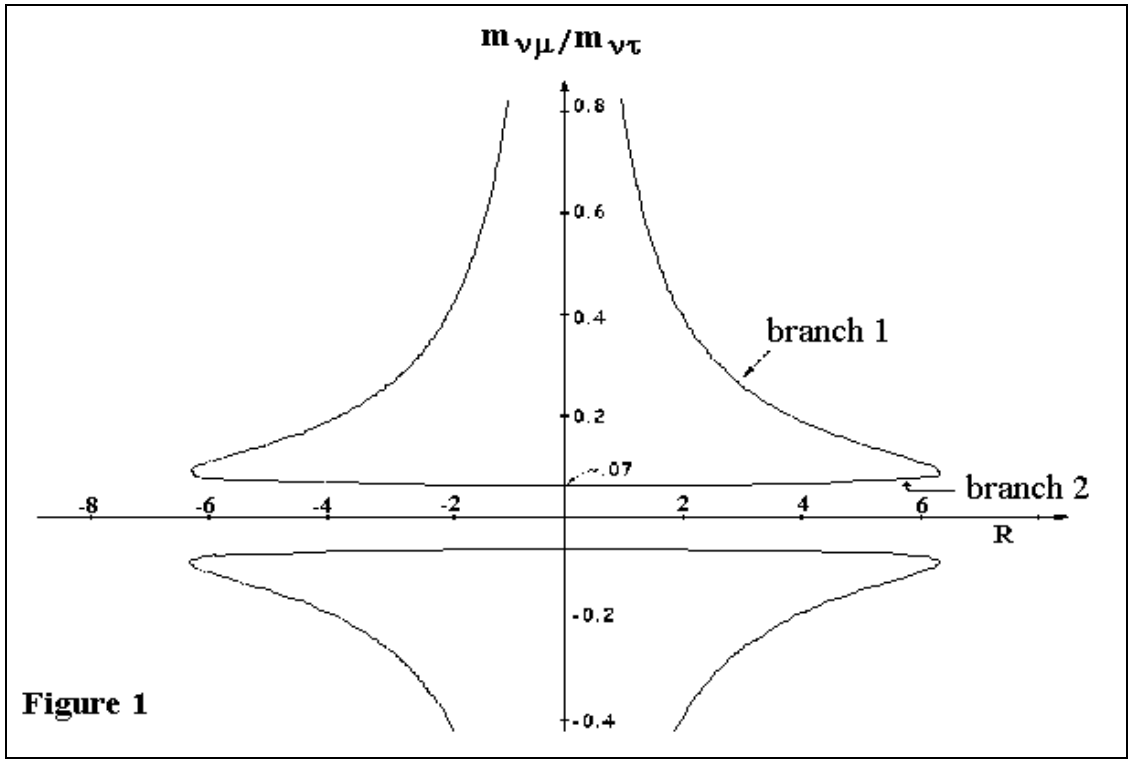

Figure 1

1. *Solution of Eqs. (2.1) and (2.2) for the ratio $m_{\nu\mu}/m_{\nu\tau}$ as a function of R. Two branches are evident, labeled branch 1 and branch 2. For branch 2 the value of $m_{\nu\mu}/m_{\nu\tau}$ is insensitive to R.*

We first remark that according to Fig. 1 there is no value for the ratio $m_{\nu\mu}/m_{\nu\tau}$ as small as $O(\lambda^4)$. Thus the data rule out the possibility that the neutrinos behave like the up-quark sector.

From Fig. 1 we see the presence of two branches. For one branch, branch 1, the masses are essentially large, although not outside the directly known bounds of neutrinos, and *not hierarchical*. We show the masses themselves for this branch in Fig. 2.

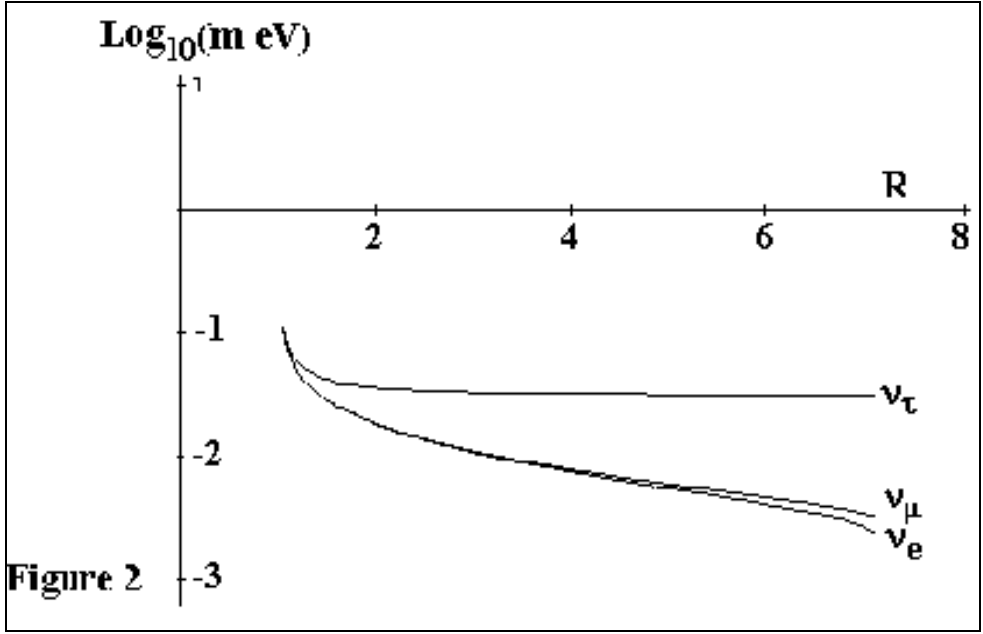

2. *Solution of Eqs. (2.1) and (2.2) for the logarithms of the masses $m_{\nu e}$, $m_{\nu \mu}$ and $m_{\nu \tau}$ as functions of R for branch 1 (see Fig. 1).*

At the minimum allowed *R*-value, which is 1, all three neutrinos masses begin essentially degenerate, and as *R* increases from its minimum allowed value the two lightest neutrino masses remain relatively close to one another. The ratio $m_{\nu\mu}/m_{\nu\tau}$ decreases somewhat but only attains a value something like $\lambda^2$ at the high end of the allowed *R* range. The second branch, branch 2 (Fig. 3), shows two features interesting to us. Firstly, the masses are arranged in a hierarchy, and secondly, the solutions for $m_{\nu\mu}$ and $m_{\nu\tau}$ are extremely stable as *R* varies, with values around

$$m_{\nu\mu} \cong 2.3 \times 10^{-3} \text{ eV and } m_{\nu\tau} \cong 0.032 \text{ eV}. \qquad (2.3)$$

The value of the ratio $m_{\nu\mu}/m_{\nu\tau}$ for branch 2 has a stable value around 0.071, which if, in line with the down quarks and the charged leptons we interpret as $\lambda^2$, corresponds to a value $\lambda = 0.27$. The fact that this is close to the measured value of the sine of the Cabibbo angle, around 0.22, as well as remarks we make at the end of this section about the other fermionic sectors, encourages us to consider branch 2 to be the relevant branch for neutrino physics.

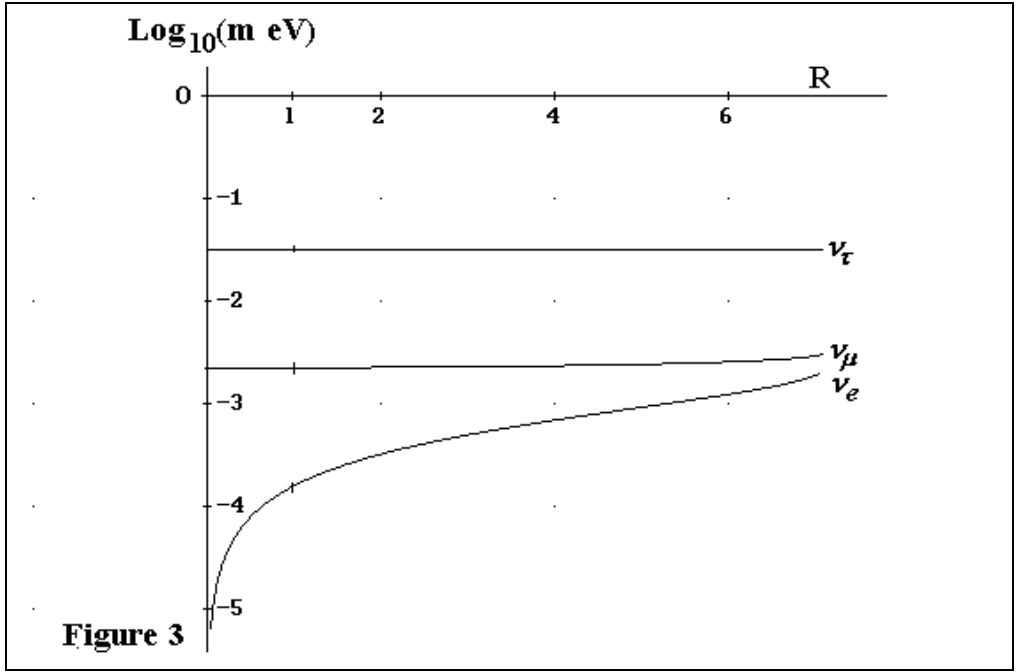

**Figure 3**

3. Solution of Eqs. (2.1) and (2.2) for the logarithms of the masses $m_{\nu e}$, $m_{\nu \mu}$ and $m_{\nu \tau}$ as functions of $R$ for branch 2 (see Fig. 1).

We are also encouraged in this view by two papers we received after the completion of this work, both of which extract the scale for the neutrino masses from the unification scale for a supersymmetric grand unified theory, very much in the spirit of the model to be discussed in the next section. Reference [12] gives $m_{\nu\tau} \cong 10^{-2}$ eV, while ref. [13] estimates $m_{\nu\tau} \cong 5 \times 10^{-2}$ eV.

While branch 2 very nearly predicts the $\mu$ and $\tau$ neutrino masses independent of $R$, the value of $m_{\nu e}$ does depends on $R$. In what follows we consider two possibilities. The case $R = 1$ corresponds to the $1:\lambda^2:\lambda^4$ pattern of the down quarks, while the case $R = .08$ ($= \lambda^2$) corresponds to the $1:\lambda^2:\lambda^6$ pattern of the charged leptons. We remark here that for the case $R = 1$, the masses for branch 2, the one of interest, take the simple form

$$m_{\nu\tau} = \frac{\Delta_{\tau\mu}}{\sqrt{\Delta_{\tau\mu} - \Delta_{\mu e}}}; m_{\nu\mu} = \frac{\sqrt{\Delta_{\tau\mu}}\sqrt{\Delta_{\mu e}}}{\sqrt{\Delta_{\tau\mu} - \Delta_{\mu e}}}; m_{\nu e} = \frac{\Delta_{\mu e}}{\sqrt{\Delta_{\tau\mu} - \Delta_{\mu e}}} \qquad (2.4)$$

While the case $R = 1$ seems to us more in line with the ideas discussed in this paper, the current data do not allow us to choose between these cases. Below we describe how the two patterns above fit into a model that generates neutrino masses. We shall treat the case $1:\lambda^2:\lambda^4$ in some detail, then add briefly how the second case, the one with a $1:\lambda^2:\lambda^6$ pattern, affects our results. We shall see below that the degree to which the electron neutrino mixes with the other two families will provide a test to distinguish between these possibilities.

It is worth reiterating here the significance of the appearance of $\lambda$ in the neutrino hierarchy. Once we believe in a particular set of hierarchies in a given fermionic sector, we can extract the values of $\lambda$ in each sector. In particular, we can take the mass ratios and hierarchy form in the different sectors as follows:

- down quarks, normalized masses run to a scale of $10^{16}$–$10^{17}$ GeV are [14] 1 to 0.034 to $1.6\times10^{-3}$ and hierarchy $1:\lambda_d^2:\lambda_d^4$.
- up quarks, normalized masses run to a scale of $10^{16}$–$10^{17}$ GeV are [14] 1 to 0.0036 to $1.3\times10^{-5}$ and hierarchy $1:\lambda_u^4:\lambda_u^8$.
- charged leptons, normalized masses are 1 to 0.059 to $2.9\times10^{-4}$ and hierarchy $1:\lambda_e^2:\lambda_e^6$.
- neutrinos, no direct data for $m_{\nu e}$, but as described above we can use $m_{\nu\mu}$ and $m_{\nu\tau}$ with a hierarchy $1:\lambda_\nu^2$.

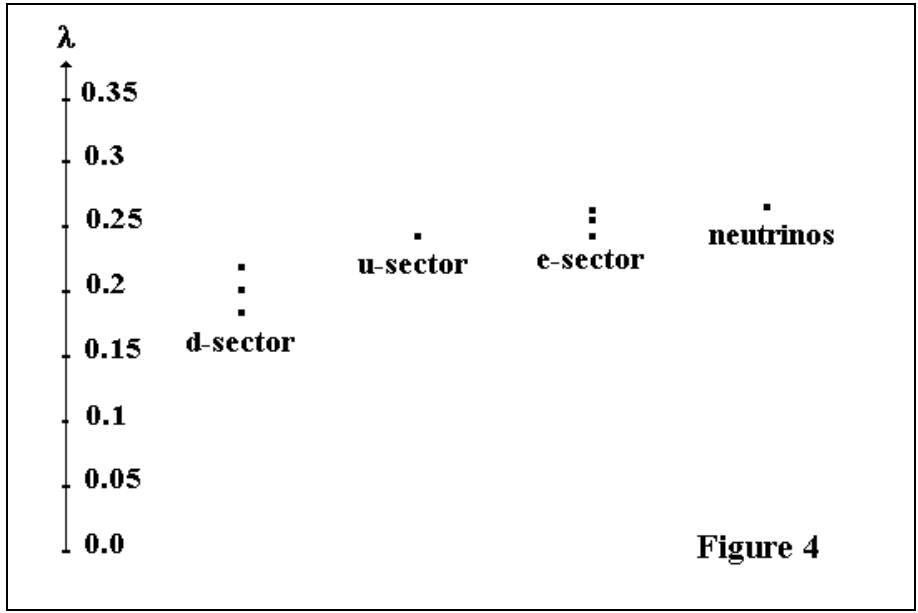

*4. The values of $\lambda$ extracted by assuming that the mass ratios in a sector are exactly equal to given powers of $\lambda$; e.g. we get $\lambda_d$ from $\lambda_d = (m_s/m_b)^{1/2}. = (0.034)^{1/2}$. Only one point is shown for the up-quarks because the same value of $\lambda$ is extracted from the three mass ratios. For the neutrinos, only the ratio $m_{\nu\mu}/m_{\nu\tau}$ with central values for these masses (see the text discussion) is used.*

Figure 4 show the values of $\lambda$ extracted by these relations, assuming a coefficient of $\lambda^n$ in the ratios exactly equal to one; *e.g.* we get $\lambda_d$ from $\lambda_d = (m_s/m_b)^{1/2}. = (0.034)^{1/2}$. We can get an idea of the possible range of the $\lambda$ found from the neutrinos by including the errors in the measured values of $\Delta_{\tau\mu}$ and $\Delta_{\mu e}$, namely (in eV$^2$) $\Delta_{\tau\mu} = 5\times10^{-4}$ to $6\times10^{-3}$ (preferred value $10^{-3}$) and $\Delta_{\mu e} = 4\times10^{-6}$ to $10^{-5}$ (preferred value $5\times10^{-6}$). We now take the naive limits and—recall the insensitivity of these masses to $R$—insert these into the $R = 1$ result [see Eq. (2.4)]

$$\lambda = \left( \frac{\sqrt{\Delta \mu e}}{\sqrt{\Delta \tau \mu}} \right)^{1/2} \qquad (2.5)$$

in order to find extreme values $\lambda = 0.16$ to $0.37$ (preferred value $0.27$). This value for $\lambda$ is certainly in the same range as the other values, around $0.25$. Perhaps, this is more than a numerical coincidence.

Finally the reader should keep in mind that even choosing a hierarchical branch, the choice of $R$, and hence of the electron neutrino mass, is not dictated by anything other than analogy to the other fermionic sectors. Here it is worthwhile to look at Figure 5. This graph allows us to understand better the sense in which the neutrinos may, or may not, interpolate the d-quark structure and the charged lepton structure.

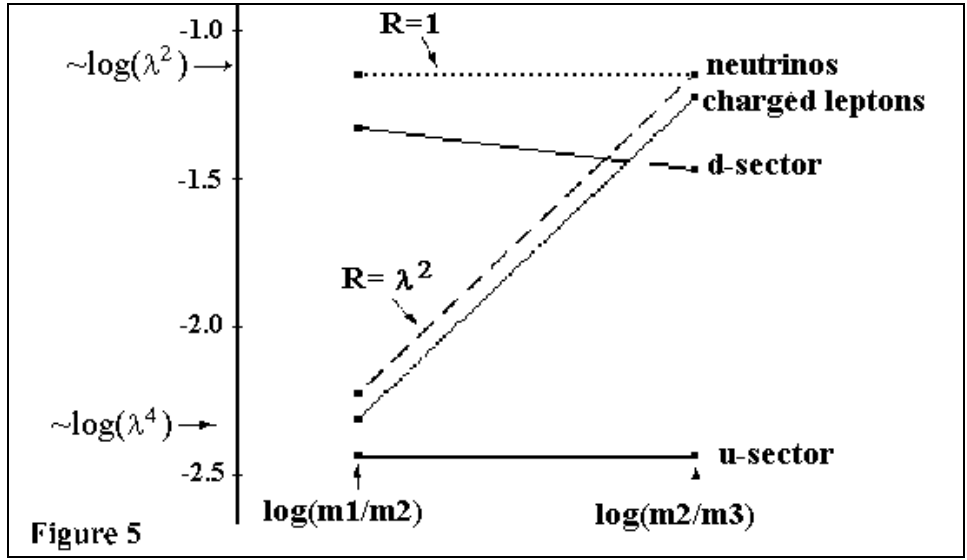

5. *The logarithms of the ratios $m_2/m_3$ and $m_1/m_2$, where 3 labels the heaviest of the three families and 1 the lightest. The solid lines connect these ratios for the three electrically charged fermionic sectors. The dotted line connects the corresponding ratios for neutrinos for $R = 1$ and the dashed line connects the corresponding ratios for neutrinos for $R = O(\lambda^2)$.*

The graph contains the mass ratios $m_2/m_1$ and $m_3/m_2$, where in each case 1 refers to the heaviest of the three families and 3 the lightest, for the three electrically charged fermionic sectors. Within both and down- and up-quark sectors these two ratios are roughly constant at $\lambda^2$ and $\lambda^4$ respectively. For the charged leptons, the first ratio is $O(\lambda^2)$ while the second is $O(\lambda^4)$. The neutrinos resemble the down quarks if $R = 1$ and the charged leptons if $R = O(\lambda^2)$.

## 3. Testing a Model of Family Symmetries

In recent work on supersymmetric models [1, 2], with additional string-inspired $U(1)$ symmetries including an anomalous $U(1)$ added to the standard model, a mechanism for the generation of charged lepton ($\ell$) and neutrino ($\nu$) masses is described. A spontaneously acquired vacuum expectation value of a supersymmetric chiral superfield gives hierarchical masses to quarks and leptons by a Frogatt-Nielsen mechanism [15]. The hierarchy is associated with charge-canceling powers in terms in the Lagrangian; connections to the string appear here, through Green-Schwarz anomaly cancellations [3], hence to determination of the charges associated with the new symmetries, and hence to the hierarchical structure of masses and mixing angles.

Here we quote the results for the mass matrices of leptons and neutrinos in these models and thereby test them. The model permits one to find, through the balancing of $U(1)$ charges, only the leading powers of $\lambda$ in the entries of the mass matrices, with unknown coefficients. In the constructive spirit of the model we assume only that these coefficients, as well as others that arise below, are of order 1. Although it is certainly possible that for unknown reasons these coefficients are as small as powers of $\lambda$ or as large as inverse powers of $\lambda$, such possibilities would ruin the predictive power and we explicitly eschew them.

*Charged leptons*. The charged lepton mass matrix takes the form [1]

$$M_\ell = \begin{bmatrix} a_{11}\lambda^4 & a_{12}\lambda^5 & a_{13}\lambda^3 \\ a_{21}\lambda & a_{22}\lambda^2 & a_{23} \\ a_{31}\lambda & a_{32}\lambda^2 & a_{33} \end{bmatrix} \qquad (3.1)$$

While the coefficients $a_{ij}$ can generally be complex, for our purposes it is enough to take them to be real, and we do so henceforth both here and in the neutrino case. Since this matrix is not hermitian, we study the hermitian form

$$H_\ell \equiv M_\ell M_\ell^\dagger; \qquad (3.2)$$

we study this form rather than its conjugate because the unitary matrix that diagonalizes $H_\ell$ will enter into the leptonic CKM matrix. We normalize so that the largest eigenvalue is unity. A quick look at the three invariants of $H_\ell$ reveals that without conditions on the $a$'s that lead to cancellations the masses squared appear in the ratio $1:\lambda^2:\lambda^{10}$, not the desired $1:\lambda^4:\lambda^{12}$. Moreover, terms in the mass matrix that are higher order in $\lambda$ are necessary to get the desired eigenvalues. The three eigenvalues of $H_\ell$ will be to leading order denoted $m_\tau^2 = 1$, $m_\mu^2 = w_2\lambda^4$, and $m_e^2 = w_3\lambda^{12}$, respectively. Higher order terms are of course necessary to get the eigenvectors correctly.

Where do the higher order correction terms come from? Presumably they are associated with quantum corrections. This is entirely reasonable in the context of the models, where $\lambda$ is proportional to the vacuum expectation value associated with a broken symmetry. Our attitude towards these terms is to put in the minimum corrections necessary to get the mass hierarchy right. While for the charged leptons we have not written the most general terms that meet this criterion, corrections other than the ones we use will not increase the size of leading contributions to the leptonic CKM matrix.

To illustrate this point, consider the corrected mass matrix

$$M_\ell = \begin{bmatrix} a_{11}\lambda^4 & a_{12}\lambda^5 & a_{13}\lambda^3 \\ a_{21}\lambda + b_{21}\lambda^2 + c_{21}\lambda^3 & a_{22}\lambda^2 & a_{23} \\ a_{31}\lambda + b_{31}\lambda^2 + c_{31}\lambda^3 & a_{32}\lambda^2 & a_{33} \end{bmatrix} \quad (3.3)$$

Calculation of the eigenvalues of $H_\ell$ shows that one can set $b_{21}$, $b_{31}$, and $c_{31}$ (or $c_{21}$) to zero and can still accommodate a nonzero value of $w_3$, but that if $c_{21}$ and $c_{31}$ are both set to zero one cannot accommodate $w_3$. Thus we set $b_{31} = c_{31} = 0$; the parameter $b_{21}$ will then have to be zero to get the masses right. We normalize to a first eigenvalue $m_\tau^2 = 1$ to leading order in $\lambda$ by the condition

$$a_{23}^2 + a_{33}^2 = 1; \quad (3.4)$$

This suggests the trigonometric representation $\theta_\ell$

$$a_{23} = \sin(\theta_\ell), \ a_{33} = \cos(\theta_\ell), \quad (3.5)$$

a form that we'll use below when convenient. Finally we insist that the hierarchy be of the desired form by setting a minimal number of these parameters in favor of $w_2$ and $w_3$; any will do, and we choose to eliminate $a_{21}$, $a_{22}$, $a_{13}$, and $c_{21}$, leading to a mass matrix of the form

$$M_\ell = \begin{bmatrix} a_{11}\lambda^4 & a_{12}\lambda^5 & \frac{a_{11}a_{33}}{a_{31}}\lambda^3 \\ \frac{a_{23}a_{31}}{a_{33}}\lambda + \frac{a_{31}\sqrt{w_2 w_3}}{a_{33}(a_{11}a_{32} - a_{12}a_{31})}\lambda^3 & \frac{\sqrt{w_2} + a_{23}a_{32}}{a_{33}}\lambda^2 & a_{23} \\ a_{31}\lambda & a_{32}\lambda^2 & a_{33} \end{bmatrix} \quad (3.6)$$

Note the presence of the $w_3$ factor in the term associated with $c_{21}$, verifying that the presence of such a correction is essential to the correct hierarchy.

The eigenvalues $m_i^2$ can now be calculated to higher orders by using the invariants; in turn the eigenvectors can be calculated as a power series in $\lambda$. The matrix $U_\ell$ that diagonalizes $H_\ell$ in the form

$$U_\ell H_\ell U_\ell^{-1} = \begin{bmatrix} m_e^2 & 0 & 0 \\ 0 & m_\mu^2 & 0 \\ 0 & 0 & m_\tau^2 \end{bmatrix} \quad (3.7)$$

is composed of rows that are the eigenvectors of the respective eigenvalues. Here we quote the correct form to $O(\lambda^5)$; the precise value of the coefficients is of no special interest here and, except for the constant $[O(\lambda^0)]$ terms, we wait until we write the leptonic CKM matrix before we insert the relevant ones.

$$U_\ell = \begin{bmatrix} 1 & O(\lambda^3)+O(\lambda^5) & O(\lambda^3)+O(\lambda^5) \\ O(\lambda^3) & \cos(\theta_\ell)+O(\lambda^4) & -\sin(\theta_\ell)+O(\lambda^4) \\ O(\lambda^3) & \sin(\theta_\ell)+O(\lambda^4) & \cos(\theta_\ell)+O(\lambda^4) \end{bmatrix} \tag{3.8}$$

The factor $w_3$ appears only in the $O(\lambda^4)$ and $O(\lambda^5)$ coefficients.

*Neutrinos*. The leading order result for the mass matrix $M_\nu$ comes from see-sawing [16] Majorana and Dirac terms in the model of ref. [1]. Unlike the charged lepton form, it is hermitian, which evidently simplifies the calculation. It takes the form

$$M_\nu = \begin{bmatrix} n_{11}\lambda^6 & n_{12}\lambda^3 & n_{13}\lambda^3 \\ n_{12}\lambda^3 & n_{22} & n_{23} \\ n_{13}\lambda^3 & n_{23} & n_{33} \end{bmatrix} \tag{3.9}$$

We shall here make the initial assumption that the $\nu_e$ mass is $O(\lambda^4)$ times the $\nu_\tau$ mass. [Below we'll describe what happens when the ratio is $O(\lambda^6)$.] The three eigenvalues of $M_\nu$ should be to leading order $m_{\nu\tau} = 1$, $m_{\nu\mu} = x_2\lambda^2$, and $m_{\nu e} = x_3\lambda^4$, respectively, where $x_2$ and $x_3$ are $O(1)$. However, without conditions on coefficients, the masses that follow from Eq. (3.9) are rather in the ratios $1:1:\lambda^6$, ratios that are inconsistent with the data as described in Section 2. Again, both conditions on the $n_{ij}$ and higher order terms in $\lambda$ are necessary to accommodate the preferred ratio(s) $1:\lambda^2:\lambda^4$. The most general form that gives the ratio $1:\lambda^2:\lambda^4$, correct to $O(\lambda^3)$ in its corrections—which essentially means $O(\lambda^2)$ corrections in the 22, 23, 32, and 33 elements—is:

$$M_\nu = \begin{bmatrix} n_{11}\lambda^6 & M_{\nu 21} & M_{\nu 31} \\ \dfrac{n_{13}\sqrt{n_{22}}+\sqrt{-x_2 x_3}}{\sqrt{n_{33}}}\lambda^3 & n_{22}+p_{22}\lambda^2 & M_{\nu 32} \\ n_{13}\lambda^3 & \sqrt{n_{22}n_{33}}+\dfrac{p_{22}n_{33}+p_{33}n_{22}-x_2}{2\sqrt{n_{22}n_{33}}}\lambda^2 & n_{33} \end{bmatrix} \tag{3.10}$$

Here $M_{\nu ji}$ stand for matrix elements which can be read off of the elements on the lower left of this symmetric form. The $p_{ij}$ are the coefficients of $\lambda^2$ in the higher order corrections to the $O(1)$ elements, and we have eliminated the coefficients $n_{12}$, $n_{23}$ and $p_{23}$ in favor of the leading mass coefficients $x_2$ and $x_3$. In addition, the normalization that gives $m_{\nu\tau} = 1$ requires that

$$n_{22} + n_{33} = 1. \tag{3.11}$$

Again, a trigonometric representation for $n_{22}$ and $n_{33}$ will be helpful, namely

$$n_{22} = \sin^2(\theta_\nu), \; n_{33} = \cos^2(\theta_\nu), \tag{3.12}$$

The next step is to find the masses more precisely, then the corresponding eigenvectors. Having done so, we can find the diagonalizing unitary matrix $U_\nu$ for $M_\nu$, whose rows are the eigenvectors:

$$U_\nu M_\nu U_\nu^{-1} = \begin{bmatrix} m_{\nu e} & 0 & 0 \\ 0 & m_{\nu\mu} & 0 \\ 0 & 0 & m_{\nu\tau} \end{bmatrix}. \tag{3.13}$$

To $O(\lambda^5)$, the form of $U_\nu$ is

$$U_\nu = \begin{bmatrix} 1 + O(\lambda^2) + O(\lambda^4) & O(\lambda) + O(\lambda^3) + O(\lambda^5) & O(\lambda) + O(\lambda^3) + O(\lambda^5) \\ O(\lambda) + O(\lambda^3) + O(\lambda^5) & \cos(\theta_\nu) + O(\lambda^2) + O(\lambda^4) & -\sin(\theta_\nu) + O(\lambda^2) + O(\lambda^4) \\ O(\lambda^3) + O(\lambda^5) & \sin(\theta_\nu) + O(\lambda^2) + O(\lambda^4) & \cos(\theta_\nu) + O(\lambda^2) + O(\lambda^4) \end{bmatrix} \tag{3.14}$$

Again, we have been explicit only for the constant terms. We remark here that the $\nu_e$ mass parameter $x_3$ already appears in terms that are first order in $\lambda$.

*Leptonic CKM*. The leptonic mixing of the weak interactions are summarized by $V$, the leptonic analog to the CKM matrix, namely

$$V = U_\ell U_\nu^{-1}. \tag{3.15}$$

If we choose the first quadrant for both $\theta_\ell$ and $\theta_\nu$, a choice that will not change the conclusions below, then to leading order in each element $V$ is

$$V = \begin{bmatrix} 1 + O(\lambda^2) & \lambda'\sqrt{\frac{-x_3}{x_2}} + O(\lambda^3) & O(\lambda^3) + O(\lambda'^3) \\ -\lambda'\sqrt{\frac{-x_3}{x_2}}\cos(\theta_\ell - \theta_\nu) + O(\lambda^3) & \cos(\theta_\ell - \theta_\nu) + O(\lambda^2) & -\sin(\theta_\ell - \theta_\nu) + O(\lambda^2) \\ -\lambda'\sqrt{\frac{-x_3}{x_2}}\sin(\theta_\ell - \theta_\nu) + O(\lambda^3) & \sin(\theta_\ell - \theta_\nu) + O(\lambda^2) & \cos(\theta_\ell - \theta_\nu) + O(\lambda^2) \end{bmatrix} \tag{3.16}$$

We have in this formula distinguished just for convenience the contribution to *V* from the diagonalizing matrix for the neutrinos ($\lambda'$) and from the diagonalizing matrix for the charged leptons ($\lambda$). The neutrino contributions dominate. We have also not bothered to specify the O($\lambda^3$) contributions in the 13 element of *V*, since the coefficient depends on the other free parameters $n_{13}$, $a_{11}$, $a_{12}$, $a_{31}$, and $a_{32}$. All the other elements are determined by the neutrino mass factors $x_2$ and $x_3$ and by the angles $\theta_\ell$ and $\theta_\nu$.

*Neutrino masses in a* **1**:$\lambda^2$:$\lambda^6$ *pattern*. This case may be very simply abstracted from the first by the simple expedient of setting $x_3 = 0$ in Eq. (3.10). The $\nu_e$ mass is then given to leading order by

$$m_{\nu e} = \frac{n_{11} n_{33} - n_{13}^2}{n_{33}} \lambda^6, \qquad (3.17)$$

with the other masses unchanged to leading order. One could, of course, replace the parameters $n_{13}$, say, in terms of this mass. The leptonic CKM matrix is equally given by Eq. (3.14) with $x_3 = 0$, i.e., it takes the modified form $V'$,

$$V' = \begin{bmatrix} 1 + O(\lambda^2) & O(\lambda^3) & O(\lambda^3) \\ O(\lambda^3) & \cos(\theta_\ell - \theta_\nu) + O(\lambda^2) & -\sin(\theta_\ell - \theta_\nu) + O(\lambda^2) \\ O(\lambda^3) & \sin(\theta_\ell - \theta_\nu) + O(\lambda^2) & \cos(\theta_\ell - \theta_\nu) + O(\lambda^2) \end{bmatrix} \qquad (3.18)$$

While the coefficients of the O($\lambda^3$) terms are not difficult to work out, they are of no special interest here. The O(1) terms are unchanged, and of course this persists for any value of $m_{\nu e}/m_{\nu \tau}$ of order smaller than $\lambda^2$.

**4. Discussion**

Independent of model, but not of bias as to whether the neutrinos masses have a hierarchical structure, the experiments that measure $\Delta_{\tau\mu}$ and $\Delta_{\mu e}$ seem to us to go a long way towards measuring the masses $m_\tau$ and $m_\mu$. Moreover, the hierarchical structure suggested comes very close to the structure of the other three fermionic sectors, a fact worthy of attention.

Next let us look in the context of the models studied at properties of the leptonic CKM, particularly Eq. (3.14). We remark first that the leading order terms are determined by the mass parameters and the two mixing angles. There are two issues that concern us, the possibility of maximal mixing in the $\mu$-$\tau$ sector, and the degree to which the $\nu_e$ mixes with the other neutrinos.

*Maximal mixing in the $\mu$-$\tau$ sector*. Equation (3.14) in the crudest approximation suggests a zeroth order mixing between the $\mu$ and $\tau$ sectors and only small mixing of the *e* sector with the $\mu$ and $\tau$ sectors. This property, and its consistency with the Kamiokande data, has been discussed already in Refs. [1] and [2], where reference is made to maximal mixing. But is truly maximal mixing, in which all four elements in the lower right hand corner of $V$ are the same size, $1/\sqrt{2}$, possible? The spirit of the model requires us to start with matrix elements of mass matrices that have a given leading power of $\lambda$ *with coefficients of O(1)*. This means that the angles $\theta_\ell$ and $\theta_\nu$ cannot be near zero, nor near $\pi/2$. An example of "maximum" coefficients in the neutrino mass matrix is $\sin^2\theta_\nu = 1/2 = \cos^2\theta_\nu$; i.e., $\theta_\nu = \pi/4$. Similarly, we have "maximal" coefficients in the mass matrix for charged leptons for $\theta_\ell = \pi/4$. But if we make this choice, then the $\mu$-$\tau$ mixing disappears to leading order in the leptonic CKM matrix.

Or we can turn this around and start with maximal mixing in the leptonic CKM matrix, for which $\theta_\ell - \theta_\nu = \pi/4$. We use this to set $\theta_\nu = \theta_\ell - \pi/4$. The $O(\lambda^0)$ terms in the mass matrices are then $\cos\theta_\ell$, $\sin\theta_\ell$, $\cos^2(\theta_\ell - \pi/4)$ and $\sin^2(\theta_\ell - \pi/4)$. In Fig. 6 we plot these four functions; 0.25 is the largest value all four of these terms can exceed, indicating that $\theta_\ell \approx 1.3$ may be the suitable region.

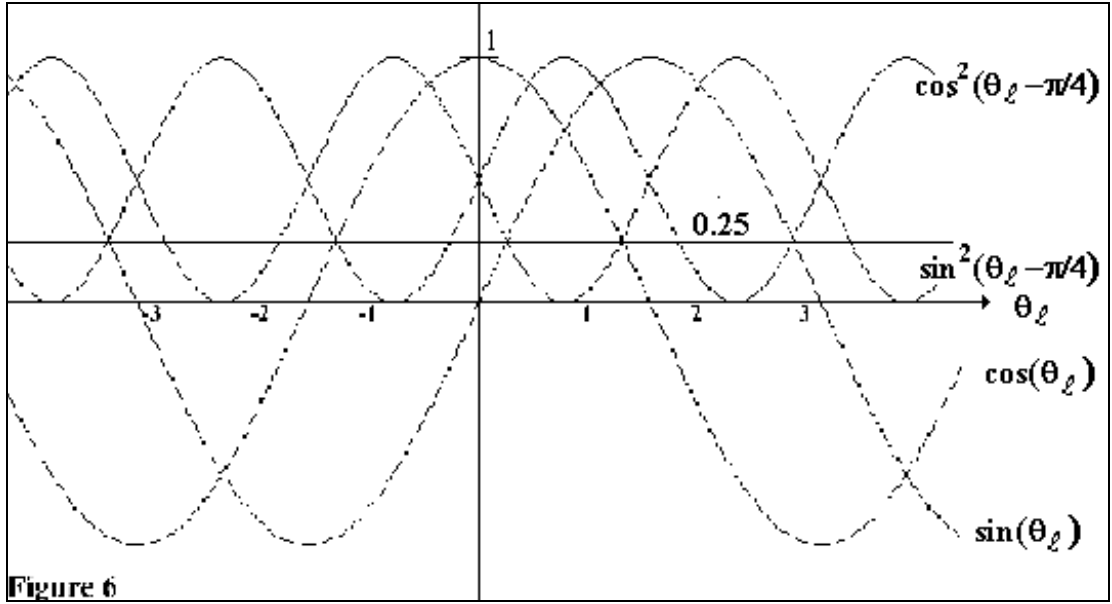

6. *Maximal mixing in the leptonic CKM is assumed, meaning $\theta_\ell - \theta_\nu = \pi/4$. We use this to set $\theta_\nu = \theta_\ell - \pi/4$. The $O(\lambda^0)$ terms in the mass matrices are then $\cos\theta_\ell$, $\sin\theta_\ell$, $\cos^2(\theta_\ell - \pi/4)$ and $\sin^2(\theta_\ell - \pi/4)$, plotted here to show the possibilities for these terms.*

*Mixing of the electron sector*. While the electron neutrino mixing is small, it does not vanish. There is a first order term in the 12, 21, and 31 elements, proportional in each case to $\lambda(-x_3/x_2)^{1/2}$. In Section 1 we argued that for $\lambda = 0.27$, $x_2 = 1$, and while we have no direct way to measure $x_2$, if $R = 1$, then $x_3 = 1$ as well. This $O(\lambda)$ term is thus just $\lambda$ itself, namely 0.27. This is not only not an unmeasurably small amount of mixing, it comes close to some of the "maximal" $\mu$-$\tau$ mixing terms for realistic values of the angles $\theta_\ell$ and $\theta_\nu$ as described above. It remains for a more careful analysis to decide whether this coupling is in fact too large to be accommodated in a small angle MSW explanation of the solar neutrino data, which may favor values of $O(\lambda^3)$.

In contrast, elements involving $\nu_e$ of $O(\lambda)$ are absent in the alternative model where neutrino masses are in a $1:\lambda^2:\lambda^6$ pattern. Here, all terms in $V$ [Eq. (3.16)] that couple the electron neutrino to the $\mu$-$\tau$ sector are $O(\lambda^3) \cong 0.02$. Thus at worst this becomes a way to distinguish the two models described here, and at best it performs a testable prediction for both versions of the model.

Acknowledgments

We thank P. Ramond for many useful conversations and the Aspen Center for Physics for its hospitality. PMF is supported in part by the U.S. Department of Energy under grant number DE-AS05-89ER40518.